# Mineral bridges in nacre revisited

**Antonio G. Checa,[1] Julyan H. E. Cartwright,[2] Marc-Georg Willinger[3,4]**

[1]Departamento de Estratigrafía y Paleontología, Facultad de Ciencias, Universidad de Granada, E-18071 Granada, Spain. [2]Instituto Andaluz de Ciencias de la Tierra, CSIC–Universidad de Granada, Campus Fuentenueva, E-18071 Granada, Spain. [3]Laboratorio Associado CICECO, Campus Universitario de Santiago, Universidade de Aveiro, 3810-193 Aveiro, Portugal. [4]Department of Inorganic Chemistry, Fritz-Haber-Institute of the Max-Planck-Society, Faradayweg 4-6, 14195 Berlin, Germany.

Abstract

We confirm with high-resolution techniques the existence of mineral bridges between superposed nacre tablets. In the towered nacre of both gastropods and the cephalopod *Nautilus* there are large bridges aligned along the tower axes, corresponding to gaps (150-200 nm) in the interlamellar membranes. Gaps are produced by the interaction of the nascent tablets with a surface membrane that covers the nacre compartment. In the terraced nacre of bivalves bridges associated with elongated gaps in the interlamellar membrane (> 100 nm) have mainly been found at or close to the edges of superposed parental tablets. To explain this placement, we hypothesize that the interlamellar membrane breaks due to differences in osmotic pressure across it when the interlamellar space below becomes reduced at an advanced stage of calcification. In no cases are the minor connections between superimposed tablets (< 60 nm), earlier reported to be mineral bridges, found to be such.

**Introduction**

Nacre is a lamellar biomaterial present today in representatives with old affinities of the molluscan classes Bivalvia, Cephalopoda, Gastropoda and Monoplacophora. It has a brick and mortar structure, in which the bricks are aragonite tablets (5-15 μm wide) and the mortar is composed of organic material (proteins and polysaccharides). Its secretion is fully extracellular and begins with the production of very closely spaced (~100 nm) parallel interlamellar membranes (Nakahara, 1983, 1991) through liquid crystallization (Cartwright and Checa, 2007, Cartwright *et al.*, 2009). These consist of a core of *β*-chitin that is surrounded by acidic proteins (Levi-Kalisman *et al.*, 2001). The intermediate spaces are filled with a gel enriched in silk fibroin. Nacre tablets begin to grow within these intermediate spaces, first quickly along the vertical direction (coincident with the aragonite *c*-axis), at the same time as the interlamellar membranes separate to the required distance to accommodate the final thickness of the aragonite plates. Tablets later expand sideways until impinging upon their neighbours, so as to fill all of the available volume.

One of the main current debates concerning nacre construction is whether the superimposed tablets are continuous or not across so called "mineral bridges" (MBs). This debate emerged with the description of nanopores (5-50 nm in diameter) in the decalcified interlamellar membranes of the gastropod *Haliotis rufescens* and the proposal that a new tablet would initiate as an offshoot of an underlying tablet through the nanopore, with no need for a new nucleation event each time a tablet is formed (Schäffer *et al.*, 1997). The mineral bridge hypothesis provided an interesting alternative to the previous heteroepitaxial hypothesis of nucleation of nacre tablets directly onto the organic scaffolding (Weiner and Traub, 1980, 1981).

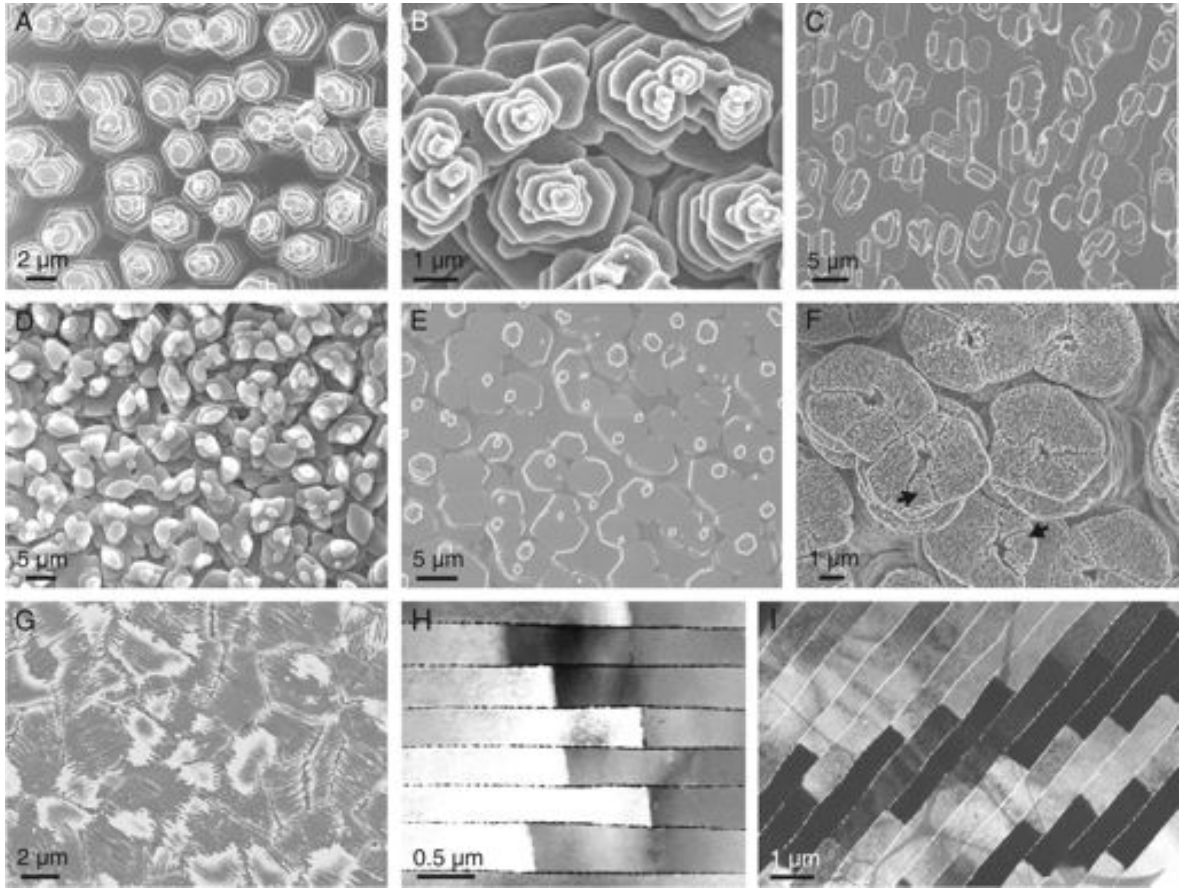

**Fig. 1.** Arrangement of nacre in molluscs. (**A**) Towered nacre in the gastropod *Osilinus lineatus*. (**B**) Marginal nacre of the internal shell surface of the cephalopod *Nautilus pompilius*. (**C**) Septal nacre of *Nautilus pompilius*. The distribution is tower-like, with up to four piled up tablets. (**D**) Marginal nacre of the bivalve *Nucula sulcata*. Tablets grow in low towers. (**E**) Internal nacre of *Nucula sulcata*. Newborn tablets always appear at the margin of parental tablets. There are two to null tablets per parental tablet. (**F**) Nacre of *Gibbula umbilicalis*. Note the general agreement in orientation between the lineations of parental and filial tablets. Some new crystals may appear (arrows) or disappear in the newborn tablets. According to lineations, polycrystalline (twinned) tablets are dominant. (**G**) Etched nacre of *Pinctada margaritifera*. The lineations revealed upon etching coincide with the crystallographic *a*-axis. Lineations of filial and parental tablets are parallel. (**H**) Diffraction contrast STEM view of *Gibbula umbilicalis* nacre. Similar greyscale intensities indicate even crystallographic alignment of tablets. (**I**) Diffraction contrast TEM photograph of *Pinctada margaritifera*. Evenly oriented tablets are arranged in a stair-like fashion. Growth direction is to the top left.

From the data available, there is little doubt that superimposed tablets communicate. Studies based on synchrotron microbeam X-ray diffraction (XRD) (Di Masi and Sarikaya, 2004), X-ray photoelectron emission spectromicroscopy (X-PEEM) (Metzler *et al.*, 2007; Gilbert *et al.*, 2008) and transmission electron microscopy (TEM) diffraction (SAED) (Gries *et al.*, 2009b) show that tablets stacked along a single column in the nacre of the abalone display similar contrast conditions, i.e., they are co-oriented. Strict co-orientation of up to ten superimposed tablets in *Mytilus edulis* has been demonstrated by means of electron backscatter diffraction (EBSD) (Dalbeck *et al.*, 2006; England *et al.*, 2007). Our own data also support this view. Both in gastropods and cephalopods nacre tablets stack in towers and the shapes of tablets of the same tower agree strictly (Fig. 1, A to C). The same applies to the terraced nacre of bivalves, in which the shape of any growing tablet always matches exactly that of the tablet immediately below (parental tablet) (Fig. 1, D and E). Etched tablets of nacre show lineations, which we know coincide with the *a*-axis of the aragonite. In the gastropod *Gibbula* plates sometimes display many sectors, with lineations intersecting at approximately either 60º or 120º (Fig. 1F). There is a good match between the lineations of superimposed tablets, although some sectors may appear or disappear from one layer to the next (arrows in Fig. 1F). In bivalves, the lineations of superimposed tablets are coincident in all those cases in which the parental-filial relation between them can be unequivocally established (Fig. 1G). Electron-diffraction reveals an agreement in crystallographic orientation in the case of gastropods (Fig. 1H) and cephalopods if the observed platelets belong to the same tower. In bivalves, an agreement in orientation is often found between tablets of different lamellae in a zig-zag manner due to the brickwall-like arrangement of tablets (Fig. 1I).

With some exceptions (Nudelman *et al.*, 2006; Suzuki *et al.*, 2009), authors have worked on the assumption that MBs are a reality and have illustrated what may be MBs in scanning (SEM) and/or TEM views (Fan and Yilong, 2001; Song *et al.*, 2002; Su *et al.*, 2002; Song and Bai, 2003; Song *et al.*, 2004; Rousseau *et al.*, 2005; Berthelat *et al.*, 2006; Velázquez-Castillo *et al.*, 2006; Lin *et al.*, 2008; Meyers *et al.*, 2008; Fengzhang *et al.*, 2009; Li and Huang, 2009; Saruwatari *et al.*, 2009; Espinosa *et al.*, 2009). Width ranges of 36-54 nm (Song *et al.*, 2004) and 25-55 nm (Gries *et al.*, 2009b) have been provided. A few authors have shown that the crystal lattice of the two superimposed tablets is continuous across the MB (Gries *et al.*, 2009b; Saruwatari *et al.*, 2009). TEM tomography in the abalone *Haliotis laevigata* has shown that what seem to be true mineral connections are but the end of a *continuum* that initiates with simple hillocks (the so-called nanoasperities, e.g., Berthelat *et al.*, 2006) that protrude from both tablets and that do not connect (Gries *et al.*, 2009b). The intra-lamellar boundaries between tablets are densely studded with purported MBs and nanoasperities (Gries *et al.*, 2009b; Fan and Yilong, 2001; Song *et al.*, 2002; Su *et al.*, 2002; Song and Bai, 2003; Song *et al.*, 2004; Lin *et al.*, 2008; Fengzhang *et al.*, 2009; Li and Huang, 2009; Saruwatari *et al.*, 2009), such that a 3D estimate from their TEM or SEM views implies that for a single tablet there would be hundreds of MBs. The only precise calculations (Song *et al.*, 2004) performed provide a figure of ~1600 MB per tablet.

On the other hand, wide connections (mean widths of 150-200 nm) have been described between the tablets of the gastropods *Gibbula* and *Monodonta* located at the axes of nacre towers (Checa *et al.*, 2009). The interlamellar membrane is totally absent here and the crystal lattice is continuous across them. The formation of the corresponding holes in the interlamellar membranes is based on the recognized existence of a surface membrane that covers the mineralization compartment of gastropod nacre and on its interaction with the incipiently growing tablets.

From the above summary, it is clear that there are two different conceptions as to what a MB should be: the widespread opinion is that all minor connections between tablets are MBs, provided that the crystal lattice is continuous across them; in the second view, MBs, at least in gastropods, are wide connections centred on the tablets. Adherents to the former view have to deal with the problem that the number of recognized MBs does not fit with the average one to one correspondence between parental and filial tablets; in this case, it has to be assumed that most of the observed MBs do not progress. In view of the present imprecision regarding the nature of MBs, we have examined the connections between tablets in representatives of the three main classes of nacreous molluscs (gastropods, bivalves and cephalopods; the rare monoplacophorans are not included owing to lack of material) by integrating data from different sources, mainly SEM and TEM. Our goal is to determine which ones of those connections are likely candidates to be true MBs, and so to present a unified picture of their formation in molluscs.

This paper is an extended version of the study we published last year (Checa et al, 2011).

**The pioneering contributions of Wada and Nakahara**

Hiroshi Nakahara, in his joint work with Bevelander (Bevelander and Nakahara, 1969) noted for the first time that the interlamellar membranes are formed before the tablets. Later, in his 1979 work (Nakahara, 1979) he described the distribution of interlamellar membranes in gastropods, together with the surface membrane. Moreover, in 1991, he described (Nakahara, 1991) the organic core in the nacre towers of gastropods and gives his model of the formation of membranes and tablets in bivalves and gastropods (Figs, 9 y 10, p. 346), which is that currently accepted.

In 1972 Koji Wada described a growth model (Wada, 1972; fig. 34, p. 158) that is clearly the precursor of the mineral bridges of Schäffer *et al.* (1997). The route from one tablet to another is via pores in the membranes that he, along with Nakahara, had recognized. Although the study is not as sophisticated as that of Schäffer *et al.*, a quarter of a century later, it is the authentic precursor of the mineral-bridges idea, and must be recognized as such. The same scheme appears reproduced in figure 10 of his work in Japanese of 1985 (Wada, 1985) and in fig. 3.19 of his book (Wada, 1999).

We dedicate this paper to Hiroshi Nakahara and Koji Wada whose pioneering work during the 1960s-1980s, begun at the National Pearl Research Laboratory in Mie, Japan, established many of the findings about nacre that our work has built upon, and whose contributions we sometimes feel are not sufficiently recognized by the community.

**Materials and Methods**

Shells of living specimens of the gastropods *Gibbula umbilicalis*, *G. pennanti*, *Monodonta* sp.*,* and *Calliostoma zyzyphinus*, the bivalve *Nucula sulcata*, and the cephalopod *Nautilus pompilius* were fixed with 2.5% glutaraldehyde in a 0.1 M cacodylate buffer. Samples were CO2 critical point dried for SEM observation. Other empty shells of the same species, as well as of the bivalves *Acila castrensis* and *Pinctada margaritifera* were also observed. Some samples were etched with 10% sodium hypochlorite for 30 mins, followed by 10 mins in a mixture of 25% glutaraldehyde and 10% acetic acid. This is a variant of Mutvei's protocol (details kindly provided by H. Mutvei, Stockholm). Other samples were bleached with commercial bleach (4% active Cl) for several minutes. In some polished sections the organic matrix was fixed with a mixture of 2.5% glutaraldehyde and 2% formaldehyde; they were further demineralized with a methanolic solution (3:1:6) in a gel medium (protocol by Hernández-Hernández et al., in prep.). All the treated shells listed above were studied using a Leo Gemini 1530 field-emission SEM (FESEM) of the Universidad de Granada.

For TEM, specimens of *G. umbilicalis*, *P. margaritifera*, *N. pompilius,* and *A. castrensis* were first mechanically polished and subsequently thinned down to electron transparency with a GATAN precision ion polishing system (PIPS) at the Fritz-Haber Institute of the Max-Planck Society in Berlin. The 3-D arrangement of nacre was studied by slice-and-view focused ion beam (FIB) tomography in the cases of the gastropod *Osilinus lineatus* and the bivalve *P. margaritifera*. Slicing (at 20 nm) and viewing was performed with an Auriga Crossbeam (FIB-SEM) workstation equipped with a Cobra ion column, belonging to Carl Zeiss Optronics GmbH (Oberkochen, Germany). The 3-D reconstruction was performed with the program ImageJ (downloadable at http://rsbweb.nih.gov/ij/). Two lamellae of *P. margaritifera* were also prepared for TEM with the Cobra ion column. TEM analysis was carried out using the Jeol 2200FS microscope at the University of Aveiro, Portugal.

**Results**

**Minor interruptions in the interlamellar membrane**

Minor protuberances or outgrowths in the size range of 10-50 nm frequently interrupt the interlamellar membranes in all species examined (Fig. 2, A to F). They are crystalline and generally terminated by flat crystalline faces (Fig. 2, A to C and F). Often, their size is large enough to bridge the gap between two tablets. In cases in which two superimposed tablets display an identical diffraction contrast, high-resolution (HRTEM) lattice fringe imaging reveals that the crystal lattice of tablets is continuous across such bridges (Fig. 2, A, B, D and E), a fact previously reported (Gries *et al.*, 2009b). However, intermediate domains with a slightly missoriented or distorted lattice sometimes appear along the crystalline connections (Fig. 1, B, D and E).

STEM views (Fig. 2, C and F) demonstrate that this crystalline continuity is illusory in many cases. The two-dimensional projection of a three-dimensional TEM lamella that contains hillock-like protrusions which may overlap at different depths with respect to the viewing direction, gives rise to the false impression of a continuous crystalline connection. STEM imaging also provides cases where an intermediate gap, which otherwise goes unnoticed in the TEM view, becomes clearly visible (Fig. 2, C, inset, and F). Some of the minor bridges in fact correspond to grains (sometimes isolated) that have grown within the interlamellar membrane (Fig. 2C). SEM observations of the surfaces of nacre tablets, i.e., the sites where minor connections may later develop, reveal the presence of small protrusions (40-100 nm wide; Fig. 1,G and H). When viewed from the top (Fig. 2I), the crystalline alignment of these nanocrystals becomes apparent.

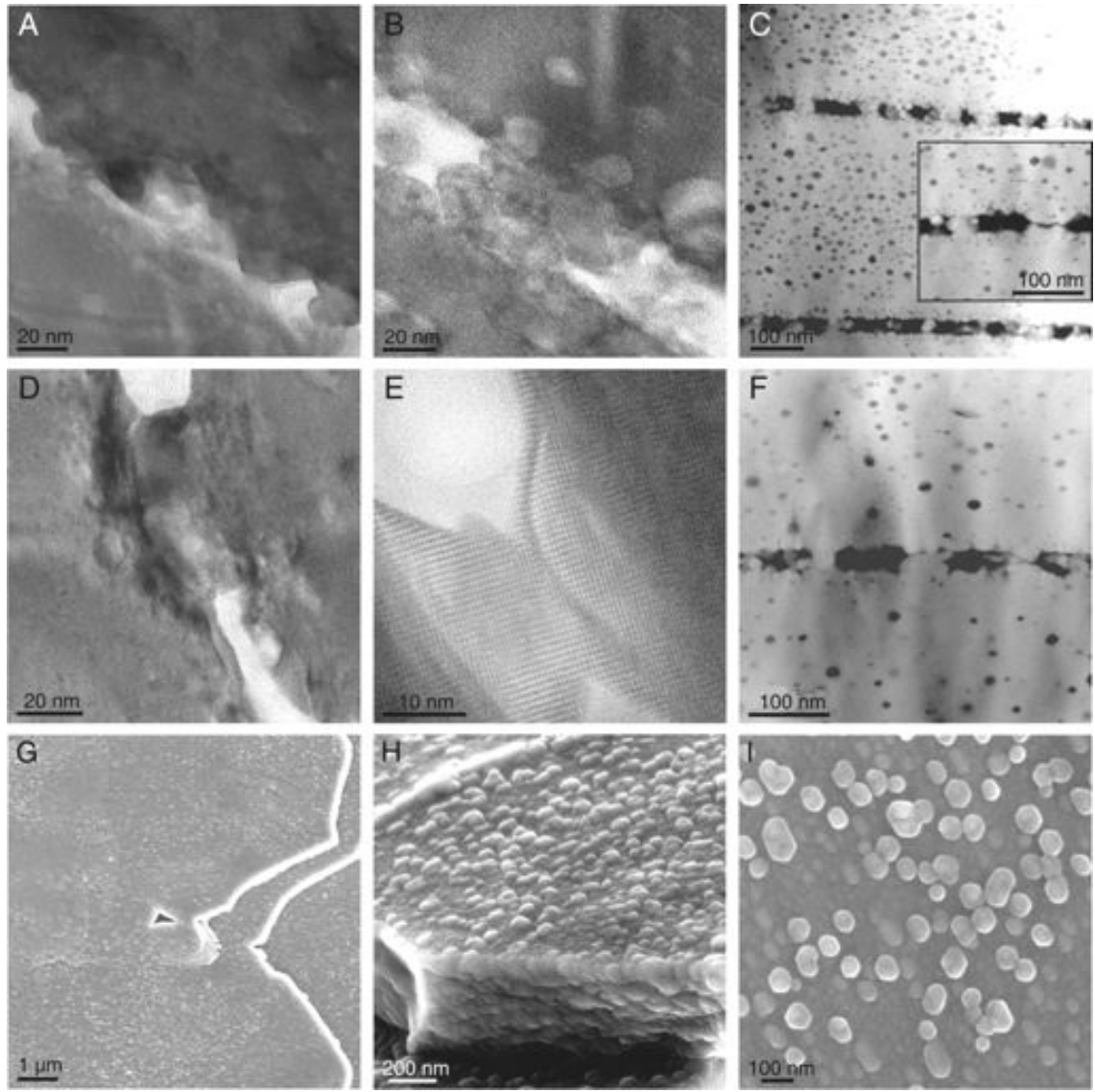

**Fig. 2.** Minor vertical connections between tablets in gastropods and bivalves. (**A-C**) *Gibbula umbilicalis*. (**D-F**) *Pinctada margaritifera*. **A**, **B**, **D** and **E** are high-resolution images of the interlamellar membranes. Despite the presence of defects, there is apparent continuity of the crystal lattice across the contacts between protrusions. **C** and **F** are STEM views of similar situations; it is clear that there is not true connection between protrusions. (**G**) Bleached internal nacre of *Nautilus pompilius*. The nanocrystalline aspect is evident. (**H-I**) Bleached nacre of *Pinctada margaritifera*. The surface is studded with nanocrystals, the aspect of which is particularly euhedral in **C**.

**Major interruptions of the interlamellar membrane**

<u>Gastropods and *Nautilus* margin</u>.- Although it was formerly reported for gastropods (Checa *et al.*, 2009), we confirm here that in both cases, the major (> 100 nm) disruptions of the interlamellar membrane appear aligned with the axes of the towers in TEM sections (Fig. 3, A and E). SEM examination of some decalcified samples of gastropods shows that these breaks are circular and appear strictly aligned with the tower axis (Fig. 3D). In TEM view, the membrane on both sides of the interruptions is often curled towards the top of the tower (Fig. 3B). In *Nautilus pompilius*, the interlamellar membrane is much less dense, with a "frayed" aspect and, sometimes, shows also additional minor breaks (Fig. 3E) in the central region of a tower. As in bivalves, HRTEM imaging and FFT analysis of the fringes demonstrate that there is crystal continuity between the two tablets in contact and across the bridge in both groups (Fig. 3, C and F).

<u>Bivalves</u>.- TEM views of *P. margaritifera*, show occasional large (> 50 nm) interruptions. When, moreover, the two tablets in contact show the same orientation, the crystal lattice is fully continuous. The aspect of two of these instances in TEM and STEM views is shown in Fig. 3, G and H. In these cases, the interlamellar membranes tend to be distorted close to the interruptions and the gap is invariably larger than 100 nm. The material at the bridge shows exactly the same aspect as in the bulk of the nacre tablets, including the typical presence of voids (Gries *et al.*, 2009a).

HRTEM and FFT imaging demonstrates that the bridge forms a crystallographic unit with the two tablets in contact (Fig. 3, G and I).

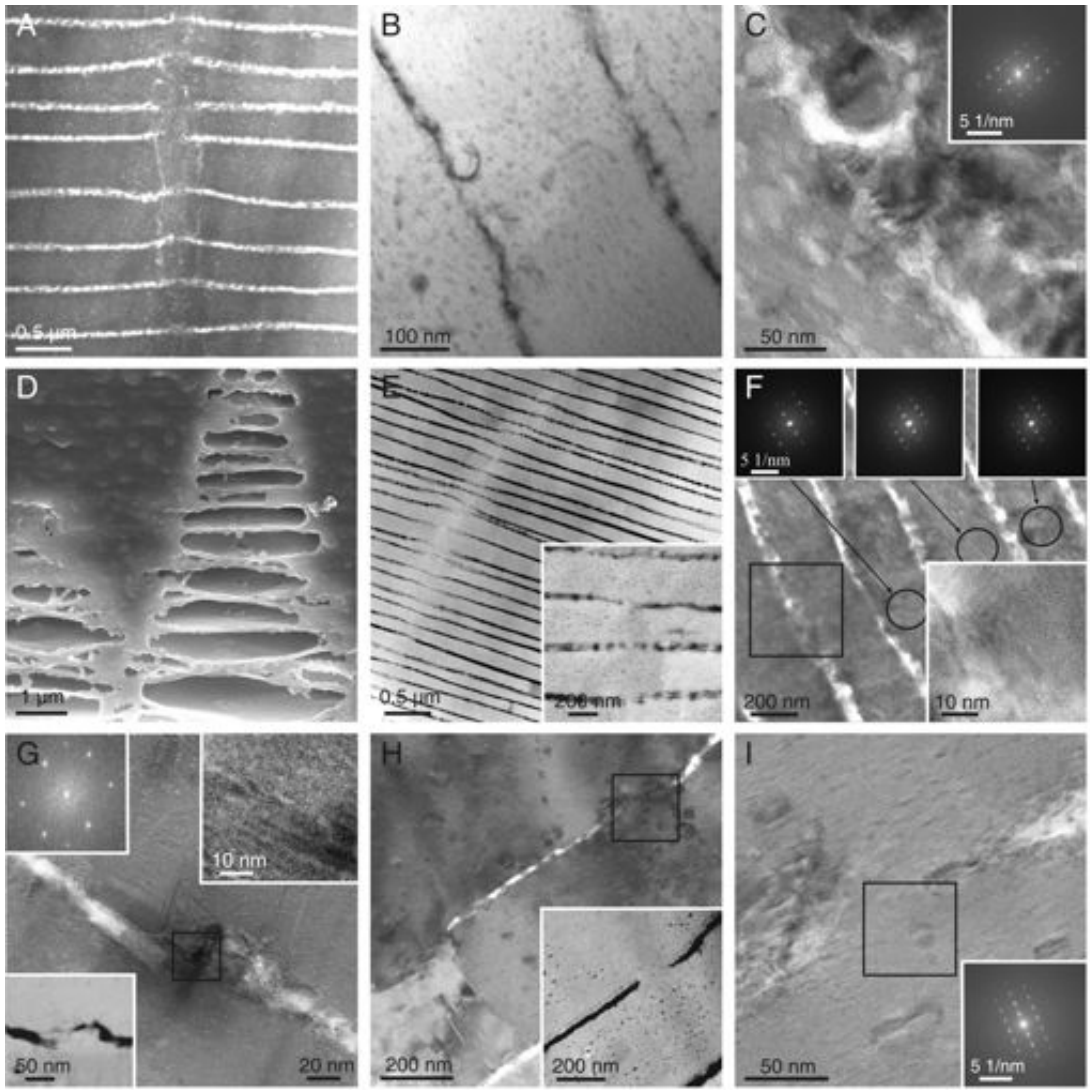

**Fig. 3.** Mineral bridges in molluscs. (**A-D**) *Gibbula umbilicalis*. (**E, F**) *Nautilus pompilius*. (**G-I**) *Pinctada margaritifera*. (**A**) TEM section through the axis of a tower displaying the interruptions of the interlamellar membranes along the tower axis. (**B**) STEM view of two similar interruptions. They are adorned on each side with upwards pointing twirls. (**C**) HRTEM view of the lower interruption in **B**. The crystal lattice is fully continuous across the interruption and the FFT (inset) indicates that the tablets above and below and the mineral bridge are one single crystal. (**D**) SEM view of a polished and decalcified sample. Note aligned holes in the interlamellar membranes. (**E**) STEM view of a nacre tower showing aligned segments of the interlamellar membranes where they become interrupted and acquire a vanished aspect. The inset is a detail. (**F**) TEM view of a mineral bridge. The HRTEM image (lower inset) and the FFTs (upper insets) imply crystal continuity across the mineral bridge (framed area). (**G-I**) Two putative mineral bridges (**G** and **H-I**). STEM views (lower insets in **G** and **H**) are also provided. The HRTEM images of the framed areas in **G** and **H** (upper right insets in **G** and **I**) and the FFTs of the framed areas in **G** and **I** (upper left inset in **G**, and lower right inset in **I**) imply both crystal unicity and continuity.

**Slice and view of *Osilinus* and *Pinctada***

*Osilinus lineatus*.- The sliced towered nacre tablets show scattered minor connections across the interlamellar membranes throughout as well as wide connections at their centres, which delineate the tower axis (Fig. 4A). Around the connections there are particularly wide empty areas, which in the 3D reconstruction of the interlamellar membranes (Fig. 4, A and B) appear as organic necks around a central hole. Evaporation of the organic phase during the milling process may both enlarge the organic neck and reduce the central hole.

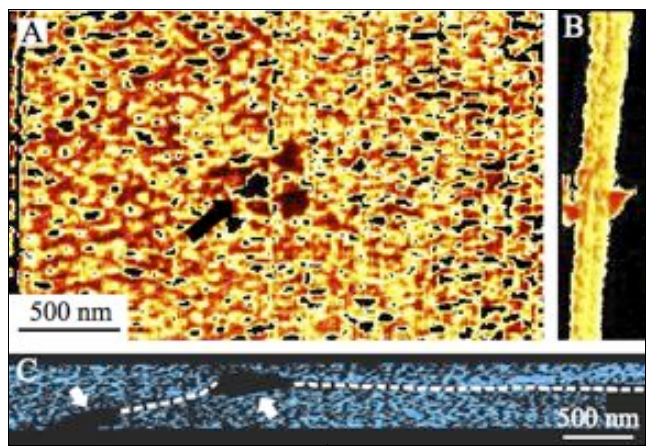

**Fig. 4.** 3D reconstructions of the interlamellar membranes in the gastropod *Osilinus lineatus* (**A-B**) and in the bivalve *Pinctada margaritifera* (**C**). (**A**) Plan view, the hole serving as passage for the mineral bridge is marked with an arrow; other minor holes scattered on the membrane correspond to connecting nanoasperities. (**B**) Side view, note organic neck around the hole. (**C**) Plan view; there are two putative passages indicated with arrows. The broken lines mark the boundaries between tablets.

*Pinctada margaritifera*.- Slicing and the associated 3D reconstruction of interlamellar membranes (Fig. 4C) show that, besides many nm-sized connections there are occasional wide connections between superimposed tablets. They have oval shapes, are several hundreds of nm wide (~ 150 x 400 nm in the examples shown in Fig. 4C), and occur preferentially at the lateral boundaries between tablets.

**Connections between tablets in gastropods and *Nautilus***

The number of minor connections between two superimposed tablets that strictly coincide in orientation, even if we consider only those that fulfil the continuous-lattice condition, would be at least several tens, a too-high figure taking into account that each parental tablet originates on average only one tablet. Even if tablets are polycrystalline, the number of sectors of the filial tablet may vary within narrow limits, but remains statistically constant. This implies that the attribution of the minor connections as true MBs is more than doubtful. This fact, together with the continuous gradation from unconnected hillocks, to hillocks that overlap in the third dimension and are connected only in appearance, to fully connected (apparent MBs), imply that these last cases are obtained when hillocks from both tablets protrude and adapt so tightly that the resulting feature appears to be continuous. This is evident in STEM view (Fig. 2C). The so-called hillocks or nanoasperities are merely nanocrystalline outgrowths or nanocrystals growing on the surfaces of the platelets, of the kind shown in Fig. 2, G to I. The latter probably grow within the proteinaceous part of the interlamellar membranes, where they have space enough to develop crystalline faces.

Major connections were recognized and described in gastropods (Checa *et al.*, 2009). We have here confirmed their nature and alignment not only in gastropods, but also in the towered nacre of the cephalopod *Nautilus*. In this group, the interlamellar membrane may either be lacking, or, alternatively, present a frayed aspect, as if it had had insufficient time to organize completely. The nature of major connections as true MBs also fits in with their position at the very centre of the tablets. The model for their origin in gastropods is based on the fact that the tips of the early nuclei of tablets grow partly embedded within the relatively thick (~100 nm) surface membrane (Checa *et al.*, 2009) (Fig. 5A). This prevents the interlamellar membrane from extending through these sites, and hence there is crystal continuity. SEM observations reveal the presence of a similar membrane in immature specimens of *Nautilus pompilius* (Fig. 5, B and C). As in gastropods, it covers the nacre compartment (Fig. 5, B, inset, and C). Its thickness in SEM view is some 100-200 nm; we still lack details as to its ultrastructure and relationship to the interlamellar membranes.

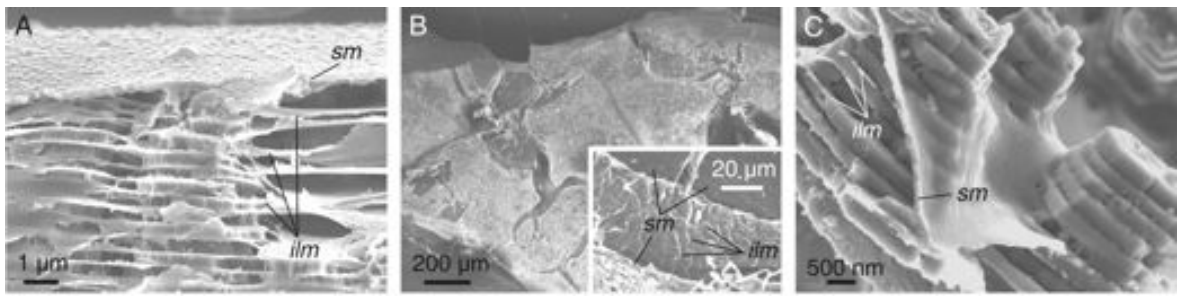

**Fig. 5.** Surface membrane in gastropods and *Nautilus*. (**A**) Fracture (side) view of the nacre compartment of the gastropod *Gibbula pennanti*, which is covered by the vesiculate surface membrane. (**B**) Surface view of the nacre close to the aperture of *Nautilus pompilius*. The surface membrane has cracked upon dehydration. The inset shows one fracture, which allows us to see the interlamellar membranes below the surface membrane. (**C**) Side view of the marginal nacre of *Nautilus pompilius*. Upon dehydration, the surface membrane has collapsed and adhered to the towers of nacre. Note the difference in thickness between the surface membrane and the interlamellar membranes. *sm* = surface membrane; *ilm* = interlamellar membrane.

The only difference between gastropods and *Nautilus* is that interlamellar membranes are not totally absent at the tower axis in the latter. Indeed, they are strongly perforated in that region. Nevertheless, the similarity in tablet distribution and the existence of a surface membrane analogous to that in gastropods (Fig. 5, B and C) imply that the mode of formation of mineral bridges in *Nautilus* is similar to that in gastropods, although the relationship of the surface membrane to the nascent tablets and interlamellar membranes must differ.

### Connections between tablets in bivalves

As in the former groups, minor and major connections have been recognized, with minor connections having exactly the same aspect and distribution. Each bivalve nacre tablet has from zero to two, very rarely three, new filial tablets. The same reasons as to the aspect, density and distribution of minor connections adduced in gastropods and *Nautilus* to discard them as real MBs apply here.

Despite their elusive nature, major interruptions in the interlamellar membrane of bivalves have been recognized in the 3D reconstructions and observed directly with TEM. The aspect of the material through the MB is exactly the same as in the tablet interior and the crystal lattice is fully continuous across the MB, without any indication of crystalline defects. The membrane has the aspect of having been disrupted at the place where the bridge formed. Such interruptions are perfect candidates for being crystalline connections between tablets, particularly since they appear close to the margins of the parental tablets and at the centres of the filial tablets, where, the continuation from below serves as a core for a newly forming tablet, thus explaining exactly what is observed with SEM and TEM (Fig. 1, E and I).

Any mineral-bridge-based explanation for the genesis of bivalve tablets has to take into account this fact that filial tablets grow from the edges of the parent tablet below, i.e., pores in the interlamellar membrane occur most often near the borders of tablets, exactly as in our 3D reconstruction. This implies that the production of new tablets happens when the parental tablet is at an advanced stage of growth or, in other words, when the volume of not yet mineralized material enclosed by the growing tablets of one layer shrinks. We hypothesize that in these cases pore formation takes place by rupture of the interlamellar membrane owing to osmotic pressure. There begins to be an osmotic pressure difference across an interlamellar membrane when tablet growth closes the compartment below the membrane, which acts as a lid on the compartment, while the zone above the membrane remains open to the general extrapallial space (Fig. 6, A and B). The composition of the extrapallial liquid in the lidded compartment alters as calcium carbonate crystallization proceeds with organic compounds becoming ever more concentrated as compartment volume diminishes. The membrane is almost certainly semipermeable, with small molecules such as water passing across it, while the larger molecules are unable to cross. Thus the increase in the concentration of organic compounds implies a corresponding increase in the osmotic pressure exerted on the lower side of the interlamellar membrane. This imbalance across the membrane leads to its bowing upwards towards the edge of tablets (Fig. 6B1), as is noted in slices through bivalve nacre (Fig. 7), and eventually to its rupture at a weak point (Fig. 6B2). Immediately after the osmotic pressure is released, the burst membrane bends into the open void, and a valley is usually observed at the very junction between tablets in areas supposedly close to the mineral bridges (Fig.

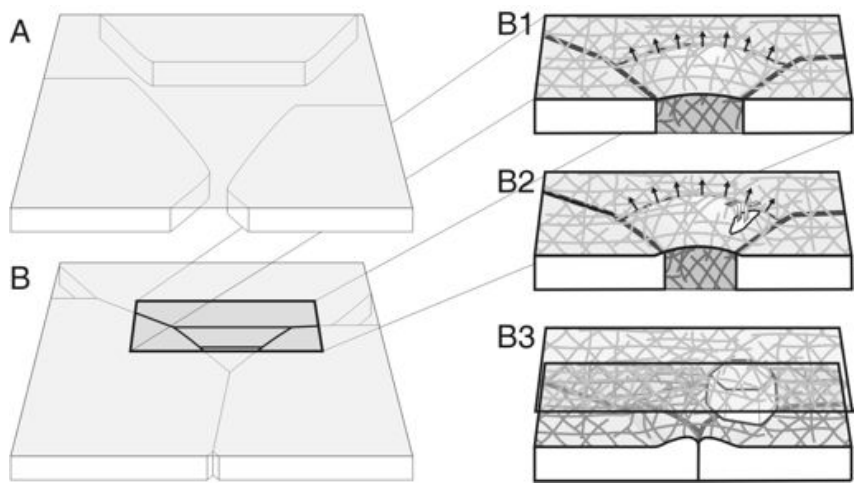

**Fig. 6.** Model for the formation of mineral bridges in bivalves. (**A**, **B**) Neighbouring tablets grow until a closed space, i.e., one with no connection to the rest of the interlamellar space, is formed. (**B1-B3**) Close up view of the processes occurring in this space. Increasing concentration of organic molecules raises the osmotic pressure within the lidded compartment and causes the interlamellar membrane to bow upwards (**B1**),

until a break is produced (**B2**). Once the osmotic pressure is released, the interlamellar membrane relaxes and a new tablet emerges from the edge of the tablet which first reaches the tear (**B3**).

6B3, and Fig. 7, arrows). This mechanism explains why pores form preferentially at the edges of tablets and also fits in with the elongated shape of the reconstructed major pores in the membranes (Fig. 4C).

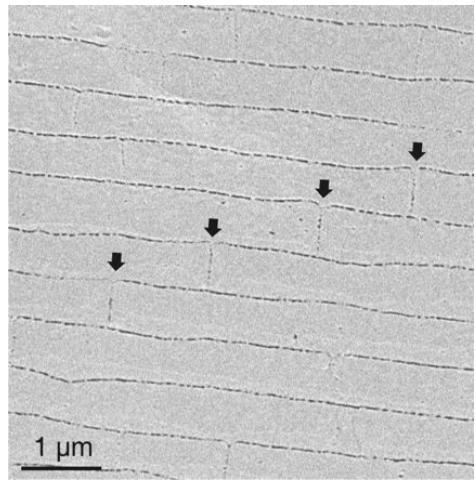

**Fig. 7.** SEM view of the polished nacre of *Pinctada margaritifera*. The interlamellar membranes are convex upwards close to the lateral contacts between tablets and the incised small valley at the contacts (arrows). Growth direction is to the top.

**Discussion**

Our study, together with former evidence (Di Masi and Sarikaya, 2004; Metzler *et al*., 2007; Gilbert *et al*., 2008; Gries *et al*., 2009b, Dalbeck *et al*., 2006; England *et al*., 2007) leads us to conclude that there is an exact match in orientation between a given tablet and the one(s) that grow(s) on top of it. This feature evidently calls for the necessary communications of aragonite crystals across the interlamellar membrane.

Nacre tablets in gastropods (Checa *et al*., 2009) and *Nautilus* communicate across circular interruptions in the interlamellar membrane, which are persistently found aligned along the axes of the towers. In both groups, the formation of axial interruptions is related to the existence of a surface membrane, although the details as to the formation of MBs may differ slightly between the groups. In bivalves, interruptions are placed close to the edges of the parental tablets. In this case, we assume that their origin is secondary, due to differences in osmotic pressure, which we hypothesize to cause bursting of the interlamellar membrane in the last stages of growth of the underlying tablets. Differences in osmotic pressure across the membrane in gastropods, if present, are so in an attenuated manner. Firstly, both the horizontal and vertical gradients in tablet growth are much lower in gastropods than in bivalves. In the latter group, fully grown tablets change into nuclei in a single lamella over a distance of several tens of microns, whereas in gastropods, tablets in a given lamella are in a similar growth state. In the vertical direction, initial nuclei of bivalves sit on top of fully-grown or almost-fully-grown tablets; this is not the case in gastropods, particularly when towers have a high aspect ratio. Finally, as the membranes of gastropods appear porous (Schäffer *et al.,* 1997; Cartwright and Checa, 2007), while those of bivalves do not (Cartwright and Checa, 2007; Cartwright *et al.*, 2009), we hypothesize that the former may be transparent to organic molecules. In any case, the number and distribution of the described major breaks exactly match those of filial tablets that emerge once a new lamella is being formed. This is not the case with other described minor breaks (<50 nm), which have been considered to be true MBs (Gries *et al.*, 2009b). On this basis, as well as on that of their morphology and distribution, we disregard them as such.

Our model implies that in nacre there is full continuity of aragonite crystals across the interruptions in the membranes; any kind of heteroepitaxy being therefore excluded. Hence, since there is no need for a nucleation event every time a new tablet is to be formed, we consider the models based on the recognition of aragonite nucleating proteins and their role in the determination of the crystallographic orientations (Suzuki *et al.*; 2009, Kröger, 2009) as superfluous.

Our study in no way exhausts the issue of how nacre grows, but rather aims at providing a new perspective for future studies. For example, there can be little hope for any effort aiming to mimic nacre in laboratory that does not take into account the increasing body of knowledge on this fascinating natural product.

**Acknowledgements**

We acknowledge Achim Klein (Fritz-Haber Institut, Berlin) and Fabian Pérez-Willard (Zeiss Optronics GmbH, Oberkochen, Germany) for sample preparation and analysis. Funding was provided by Projects CGL2007-60549, CGL2010-20748-CO2-01 and FIS2010-22322-C02-02 of the Spanish Ministerio de Ciencia e Innovación, Projects REDE/1509/RME/2005 and PTDC/CTM/100468/2008 of the Portuguese Foundation for Science and Technology (FCT), the Research Group RNM190 (Junta de Andalucía), and the European COST Action TD0903.